\documentclass[11pt,a4paper,reqno]{amsart}%
\usepackage[latin1]{inputenc}
\usepackage{mathrsfs}
\usepackage{dsfont}
\usepackage{hyperref}
\usepackage{amsmath}
\usepackage{amssymb}
\usepackage{amsthm}
\usepackage{amsfonts}
\usepackage{amstext}
\usepackage{amsopn}
\usepackage{amsxtra}
\usepackage{mathrsfs}
\usepackage{dsfont}
\usepackage{graphicx}
\usepackage{color}

\newtheorem{theorem}{Theorem}[section]

\newtheorem{lemma}[theorem]{Lemma}

\numberwithin{equation}{section}

\newcommand{\ii}{\infty}
\newcommand\R{{\ensuremath {\mathbb R} }}
\newcommand\C{{\ensuremath {\mathbb C} }}
\newcommand\N{{\ensuremath {\mathbb N} }}

\newcommand\1{{\ensuremath {\mathds 1} }}

\renewcommand\phi{\varphi}

\newcommand{\gH}{\mathfrak{H}}

\newcommand{\wto}{\rightharpoonup}

\newcommand{\cE}{\mathcal{E}}

\newcommand{\cF}{\mathcal{F}}

\newcommand{\eps}{\epsilon}

\newcommand{\F}{\mathcal{F}}

\renewcommand{\epsilon}{\varepsilon}

\newcommand{\norm}[1]{ \left| \! \left| #1 \right| \! \right| }
\DeclareMathOperator{\tr}{{\rm Tr}}

\renewcommand{\ge}{\geqslant}

\renewcommand{\geq}{\geqslant}
\renewcommand{\leq}{\leqslant}

\newcommand{\FNL}{F_{\mathrm{NL}}}
\newcommand{\Om}{\Omega}
\newcommand{\NN}{\mathcal{N}}

\title[]{From bosonic grand-canonical ensembles to nonlinear Gibbs measures}
\author[N. Rougerie]{Nicolas ROUGERIE}
\address{Universit\'e Grenoble 1 \& CNRS,  LPMMC (UMR 5493), B.P. 166, F-38042 Grenoble, France}
\email{nicolas.rougerie@grenoble.cnrs.fr}

\date{July 2015}

\begin{document}

\begin{abstract}
In a recent paper, in collaboration with Mathieu Lewin and Phan Th\`anh Nam, we showed that nonlinear Gibbs measures based on Gross-Pitaevskii like functionals could be derived from many-body quantum mechanics, in a mean-field limit. This text summarizes these findings. It focuses on the simplest, but most physically relevant, case we could treat so far, namely that of the defocusing cubic NLS functional on a 1D interval. The measure obtained in the limit, which lives over $H^{1/2-\eps}$, has been previously shown to be invariant under the NLS flow by Bourgain.
\end{abstract}

\maketitle

\setcounter{tocdepth}{2}
\tableofcontents

%%%%%%%%%%%%%%%%%%%%%%%%%%%%%%%%%%%%%%%%%%
%%%%%%%%%%%%%%%%%%%%%%%%%%%%%%%%%%%%%%%%%%
\section{From many-body quantum mechanics to nonlinear Schr\"odinger theory}\label{sec:intro}
%%%%%%%%%%%%%%%%%%%%%%%%%%%%%%%%%%%%%%%%%%
%%%%%%%%%%%%%%%%%%%%%%%%%%%%%%%%%%%%%%%%%%

The nonlinear Schr\"odinger equation is a basic model of quantum optics and condensed matter physics, whose analysis has motivated a huge mathematical literature. The equation itself has been originally introduced as a phenomenological model, and its derivation from the fundamental axioms of quantum mechanics is a major issue in theoretical and mathematical physics. 

The advent of Bose-Einstein condensates of ultracold dilute alkali vapors in the mid-90's has given a renewed impetus to the field. The NLS equation is nowadays thoroughly studied as the basic effective model for the description of such objects. The reason why a one-body model such as NLS theory can be used to describe a many-body system is that, in this context, all particles reside in the same quantum state (this is what is meant by Bose-Einstein condensation).

An important line of research has been the rigorous derivation of the equation, in a certain limit, from the many-body Schr\"odinger equation for bosonic particles. This section recalls some background on this topic. The author apologizes for not attempting to do justice to the relevant literature in any exhaustive way: the reader is refered to~\cite{BenPorSch-15,Golse-13,LieSeiSolYng-05,Rougerie-LMU,Rougerie-cdf,Schlein-08} for a more complete introduction to the subject.

\subsection{Derivation of ground states}\label{sec:GS}

The basic object one starts from is a many-body Schr\"odinger Hamiltonian, say of the form
\begin{equation}\label{eq:H_N}
H_N = \sum_{j=1} ^N \left( -\Delta_j + V (x_j) \right) + \sum_{1\leq i<j\leq N} w_N (x_i - x_j).
\end{equation}
Here $V$ is an external potential felt by the particles and $w_N$ a pair-interaction potential, that is most often taken radial. This operator is considered as acting on $L_s ^2 (\R ^{dN}) \simeq \bigotimes_s ^N L ^2 (\R ^d)$, the space of square integrable, symmetric $N$-body bosonic wave functions, that is of those $\Psi_N \in L ^2 (\R ^{dN}) \simeq \bigotimes ^N L ^2 (\R ^d)$ satisfying 
$$ \Psi_N (x_1,\ldots,x_N) = \Psi_N (x_{\sigma(1)},\ldots,x_{\sigma(N)})$$
for all permutation $\sigma$ of the coordinates $x_1,\ldots,x_N \in \R ^d$ of the particles under consideration. This symmetry requirement means that we consider bosonic particles, which, as the name indicates, is mandatory to obtain a Bose-Einstein condensate.  On a physical level this implies that we consider the atoms of the gas as the basic particles, ignoring that they are themselves made of more fundamental objects. This is however as fundamental as one can get for a reasonble description. 

By ground state we mean the state of lowest possible energy, i.e. the eigenfunction of the operator achieving the lowest eigenvalue, or, equivalently, solving the minimization problem for the quadratic form
\begin{equation}\label{eq:ener N}
\cE_N [\Psi_N] = \langle \Psi_N, H_N \Psi_N \rangle  
\end{equation}
under the constraint that $\Psi_N \in L ^2_s (\R ^{dN})$ be $L^2$-normalized. This gives the most stable equilibrium configuration at zero temperature.

An important question is the following: is it true that, at least in some well-defined scaling limit, the ground state(s) of the above Hamiltonian resembles a Bose-Einstein condensate?  The latter is a $N$-body wave function of the form
\begin{equation}\label{eq:BEC ansatz}
 \Psi_N (x_1,\ldots,x_N) = \prod_{j=1} ^N u(x_j), 
\end{equation}
corresponding to putting all the particles in the same quantum state $u\in L^2 (\R ^d)$. In classical mechanics this would correspond to assuming independent identically distributed particles, i.e. molecular chaos, Boltzmann's Stosszahlansatz.

If such an ansatz is valid, one naturally obtains a NLS-like energy functional that $u$ should minimize. Indeed, inserting in~\eqref{eq:ener N} we find
$$ \cE_N [u ^{\otimes N}] = N \left( \int_{\R ^d} |\nabla u | ^2 + V |u| ^2 \right) + \frac{N(N-1)}{2} \iint_{\R ^d \times \R ^d} |u(x)| ^2 w_N (x-y) |u(y)| ^2 dx dy. $$
One usually expects the ansatz~\eqref{eq:BEC ansatz} to be valid for large particle number $N$. In order that the one- and two-body terms in the above functional weigh the same in this limit, so that one might hope for a well-defined limit, it is natural to assume that $w_N$ be of order $N-1$. At least two possibilities to achieve this should be considered:

\medskip

\noindent$\bullet$ For a fixed potential $w$, pick
$$ w_N = \frac{\lambda}{N-1} w$$
for some $\lambda \in \R$  kept fixed (or converging to a constant) when $N\to \infty$. This situation is often called the mean-field limit, it physically corresponds to having many weak collisions between the particles. The limit functional one obtains is then
\begin{equation}\label{eq:Hartree func}
\cE [u] = \int_{\R ^d} |\nabla u | ^2 + V |u| ^2 + \frac{\lambda}{2} \iint_{\R ^d \times \R ^d} |u(x)| ^2 w (x-y) |u(y)| ^2 dx dy, 
\end{equation}
i.e. a nonlinear functional with non-local nonlinearity, often called Hartree's functional.

\medskip

\noindent$\bullet$ For a fixed constant $\beta > 0$ and a fixed potential $w$, pick 
$$ w_N = \frac{\lambda}{N-1} N ^{d\beta} w (N ^\beta . )  $$
with $\lambda$ as above. This can be called a NLS limit. It allows to take the range of the potential as a parameter, and to describe a situation where, because of diluteness, collisions can be rare but  quite strong. This happens when $\beta > 1/d$. Since 
$$ N ^{d\beta} w (N ^\beta . ) \wto \left(\int_{\R ^d } w \right) \delta_0 $$
as measures, it is natural to expect a true NLS functional with local nonlinearity
\begin{equation}\label{eq:NLS func}
\cE [u] = \int_{\R ^d} |\nabla u | ^2 + V |u| ^2  + \frac{\lambda}{2} \left(\int_{\R ^d } w \right) \int_{\R ^d} |u(x)| ^4 dx
\end{equation}
in the limit.

\medskip

We refer to the literature, in particular the previously mentioned reviews~\cite{LieSeiSolYng-05,Rougerie-LMU} and also~\cite{BenLie-83,LieYau-87,LieSei-06,FanSpoVer-80,RagWer-89} for results in this direction. Under fairly general assumptions, one may obtain convergence of the energy per particle:
$$ \frac{E(N)}{N} \to e \mbox{ when } N\to \infty$$
where $E(N)$ is the bottom of the spectrum of $H_N$ and $e$ the infimum (under a unit $L^2$-mass constraint) of the relevant limit functional. This serves as a justification for the use of NLS functionals to describe the ground states of large bosonic systems. We warn the reader that for $\beta$ relatively large in the NLS limit, important new aspects emerge and an ansatz~\eqref{eq:BEC ansatz} is no longer sufficient, see~\cite{LieSei-06,NamRouSei-15} as well as~\cite{LieSeiSolYng-05} and references therein. The most prominent case where these important subtleties occur is when $d=3$ and $\beta = 1$, the so-called Gross-Pitaevskii limit. Then, short-range correlations in the ground state have the effect of replacing $\int_{\R ^3} w$ by $4\pi a$ in~\eqref{eq:NLS func},  where $a$ is the scattering length of $w$.

One can also prove results on (sequences of) $N$-body minimizers, showing that in a certain specific sense, they are close to factorized states of the form~\eqref{eq:BEC ansatz}. Let us stay vague on this point for the moment and simply warn the reader that in general, the wave function $\Psi_N$ itself is not close to a condensate, only its reduced density matrices are.

% This requires to introduce the reduced density matrices of a state $\Psi_N$: 
% \begin{equation}\label{eq:red mat N}
% \gamma_N ^{(k)} = \tr_{k+1\to N} \left[ \right] 
% \end{equation}

\subsection{Derivation of evolution equations}\label{sec:dyn}

Once one knows that ground states of large bosonic systems show Bose-Einstein condensation, the next natural question is wether this is preserved by the many-body Schr\"odinger dynamics. Indeed, a typical experimental situation is as follows: create a Bose-Einstein condensate by cooling a Bose gas to extremely low temperatures, so that one may assume that the ground state is reached. The $N$-particles are then in a stationary state of the Hamiltonian $H_N$ which is (approximately) a factorized state $u ^{\otimes N}$. To probe the dynamics one may then perturb the system, most often by changing the one-body potential $V$ in~\eqref{eq:H_N} and observe the subsequent evolution. 

This asks the question of the dynamics of (approximately) factorized states under the quantum evolution associated to Hamiltonians of the form~\eqref{eq:H_N}. Consider the Cauchy problem
\begin{equation}\label{eq:Cauchy N corps}
\begin{cases}
i \partial_t \Psi_N  = H_N \Psi_N  \\
\Psi_N (0) \approx u_0 ^{\otimes N}
\end{cases}
\end{equation}
where we stay voluntarily vague on the precise meaning of the symbol $\approx$. Usually this means convergence of reduced density matrices, the typical result one obtains for ground states. Is it true that the solution $\Psi_N (t)$ stays approximately factorized
\begin{equation}\label{eq:dyn BEC}
 \Psi_N (t,x_1,\ldots,x_N) \approx \prod_{j=1} ^N u (t,x_j)  
\end{equation}
after some time evolution? In limits similar to those previously mentioned, this has been proved by a variety of authors. Inserting the ansatz in~\eqref{eq:Cauchy N corps} one sees that the relevant $u(t)$ solves a nonlinear equation:

\medskip 

\noindent$\bullet$ Hartree's equation 
 \begin{equation}\label{eq:Hartree eq}
\begin{cases}
i \partial_t u  = -\Delta u + V u + \lambda (w\ast |u| ^2) u\\
u(0) = u_0
\end{cases}
\end{equation}
in the case of a mean-field limit.

\medskip

\noindent$\bullet$ a NLS equation 
 \begin{equation}\label{eq:NLS eq}
\begin{cases}
i \partial_t u  = -\Delta u + V u + \left(\int_{\R ^d } w \right) \lambda |u| ^2 u\\
u(0) = u_0
\end{cases}
\end{equation}
in the case of a NLS or Gross-Pitaevskii limit. In the latter case, $\int_{\R ^d } w$ should be replaced by $4\pi \times$ (scattering length of $w$).

\medskip

Many results in this direction have been proved over the years, see the reviews~\cite{BenPorSch-15,Golse-13,Schlein-08} and for example the (by far non exhaustive) following list of original references~\cite{AmmNie-15,Hepp-74,GinVel-79,Spohn-80,ErdSchYau-09,FroKnoSch-09,RodSch-09,Pickl-11}.

\subsection{Positive temperature stationary states}\label{sec:statio}

The results quickly summarized in the previous subsections already put the NLS theory on a firm rigorous ground as far as the description of Bose-Einstein condensates at low temperature is concerned. But it would be highly desirable to know something about not only ground states as in Subsection~\ref{sec:GS} but also positive temperature states. Indeed, in experiments the temperature is always (extremely) low but finite. The Bose-Einstein condensation phenomenon in fact refers to the existence of a critical temperature below which the positive-temperature equilibrium states of a Bose gas show condensation in the sense that all particles reside in the ground state of NLS functionals as previously defined. A satisfactory, mathematically rigorous, estimate of the critical temperature is still missing. 

A closely related question is that of the description of stationary, positive temperature, states in the absence of Bose-Einstein condensation. Can the NLS description still be relevant, in a well-defined limit, for such objects? The (canonical) equilibrium state for the Hamiltonian $H_N$ at temperature $T$ is the Gibbs state 
\begin{equation}\label{eq:Gibbs canonical}
\Gamma_{N,T} = \frac{1}{Z_{N,T}} \exp\left( -\frac{1}{T} H_N \right), 
\end{equation}
a trace-class operator on $L ^2_s (\R ^{dN})$, where the partition function $Z_{N,T}$ fixes the trace to be~$1$. Can such states, or close variants, converge to NLS-related objects in some limit? What would the natural limiting objects be? These questions are the subject of the paper~\cite{LewNamRou-14d} that we summarize in this text. Here we shall first try to answer the second question. Since many-body Gibbs states are stationary under the many-body Schr\"odinger flow, one should expect that the natural limit objects are invariant under the NLS flow.

%%%%%%%%%%%%%%%%%%%%%%%%%%%%%%%%%%%%%%%%%%
%%%%%%%%%%%%%%%%%%%%%%%%%%%%%%%%%%%%%%%%%%
\section{Nonlinear Gibbs measures and the NLS flow}\label{sec:Gibbs}
%%%%%%%%%%%%%%%%%%%%%%%%%%%%%%%%%%%%%%%%%%
%%%%%%%%%%%%%%%%%%%%%%%%%%%%%%%%%%%%%%%%%%

Nonlinear Gibbs measures have recently become a useful tool to construct solutions to time-dependent nonlinear Schr\"odinger equations with rough initial data, see for instance~\cite{LebRosSpe-88,Bourgain-94,Bourgain-96,Bourgain-97,Tzvetkov-08,BurTzv-08,BurThoTzv-10,ThoTzv-10,CacSuz-14}. These are probability measures which are \emph{formally} defined by
 \begin{equation}
d\mu(u)=\frac{1}{Z}e^{-\cE[u]}\,du,
\label{eq:mu_intro}
 \end{equation}
where $\cE [u]$ is a NLS-like functional as discussed above. The rigorous construction of these measures is an important part of the Euclidean approach to Constructive Quantum Field Theory (CQFT)~\cite{DerGer-13,GliJaf-87,LorHirBet-11,Simon-74,Summers-12,VelWig-73}. The recent progress on the PDE side has mainly been to define the NLS flow almost surely on the support of the measures, which typically live over rather rough functional spaces, and to prove the invariance of the measures along the associated NLS flow.

These objects have an obvious formal resemblance with Gibbs states such as~\eqref{eq:Gibbs canonical} and it is the main goal of the paper~\cite{LewNamRou-14d} to explore this connection in more details. For the moment we can treat only the defocusing 1D case based on taking $d=1$ and $\lambda,w \geq 0$ in~\eqref{eq:Hartree func} or~\eqref{eq:NLS func}. This is due to well-known difficulties with the definition of the measures in other cases. For this reason the rest of the discussion will be mainly limited to the case $d=1$, although we have partial results in a more general and abstract setting, for which we refer to the original paper~\cite{LewNamRou-14d}.  

\subsection{A reminder on classical dynamics}\label{sec:class}

Consider the evolution of a single classical particle. The dynamics is given by a Hamilton function $H(x,p) : \R ^d \times \R ^d \to \R $ on the phase space of all possible positions and velocities. The Hamilton equation for the dynamics of the particle is 
$$ \begin{cases}
   \dot{x} = \nabla_p H(x,p)\\
   \dot{p} = -\nabla_x H(x,p). 
   \end{cases}
$$
Define the associated Gibbs measures at temperature $T$
\begin{equation}\label{eq:class Gibbs}
 d \mu (x,p) = \frac{1}{Z} \exp\left( - T ^{-1} H(x,p) \right) dx dp 
\end{equation}
where $Z$ is a normalization factor ensuring
$$ \iint_{\R ^d \times \R ^d }d \mu (x,p) = 1.$$
These measures are of course invariant under the flow, because the latter conserves the energy $H(x,p)$ and leaves the Lebesgue measure invariant (Liouville's theorem). One should note that the Gibbs measures are not \emph{any} invariant measure, in that they have a physically very natural variational characterization. Namely, they minimize the free-energy functional 
$$ \iint_{\R ^d \times \R ^d } H(x,p) d\mu (x,p) + T \iint_{\R ^d \times \R ^d } \mu \log \mu \: dxdp$$
amongst probability measures, which means that they achieve a balance between minimizing the energy and maximizing the entropy.

Again by formal analogy, between~\eqref{eq:mu_intro} and~\eqref{eq:class Gibbs} this time, the measures~\eqref{eq:mu_intro} are good candidates to play a similar role in the NLS theory where the dynamics is (at least formally) also Hamiltonian.

\subsection{Invariant measures for the NLS equation}\label{sec:NLS flow}

Let us now recall the rigorous definition of the Gibbs measure~\eqref{eq:mu_intro}, based on the functionals~\eqref{eq:Hartree func} or~\eqref{eq:NLS func}. The way to proceed is by first defining the free, non-interacting, measure (case $\lambda = 0$), formally
$$ \mu_0 (du) = \frac{1}{Z_0}\exp\left( - \int_{\Om} |\nabla u | ^2 \right) du$$
where $\Om \subset \R ^d$. Then one defines the interacting measure as being absolutely continuous with respect to $\mu_0$: For a given function $w$ or even for $w=\delta_0$, set
\begin{equation}\label{eq:mu rig}
\mu (du) = \frac{1}{Z_r} \exp\left( - \frac{1}{2}\iint_{\Om\times\Om} |u(x)| ^2 w(x-y) |u(y)| ^2 \right) \mu_0 (du) 
\end{equation}
where the number $Z_r$ (relative partition function) ensures that this is a probability measure. In~\cite{LewNamRou-14d} we have considered only cases where this construction makes sense without any need for renormalization. 

\subsubsection*{The free measure} The free measure is a gaussian measure over an infinite dimensional Hilbert space. It is defined by the following classical construction~\cite{Bogachev,Skorokhod-74}. Take $h\geq 0$ a self-adjoint operator on some Hilbert space $\gH$ with compact resolvent. For the applications in this note we shall take
\begin{equation}\label{eq:anharm osc}
h = -\frac{d ^2}{dx ^2} + |x| ^a + m  
\end{equation}
on $\R$, or 
\begin{equation}\label{eq:Lap inter}
h = -\frac{d ^2}{dx ^2} + m 
\end{equation}
on $[-1,1] \subset \R$ with either, Dirichlet, Neumann or periodic boundary conditions. The $L ^2$ space on which these operators act will always be denoted $\gH = L ^2 (\Omega)$. We require that $m\in \R$ is such that these operators are positive definite. In the latter two cases one must pick $m>0$ to ensure this. We shall restrict to $a>2$ for reasons that we will explain later, and include the case~\eqref{eq:Lap inter} as a formal limit $a\to \infty$ of~\eqref{eq:anharm osc}.

We write the spectral decomposition\footnote{$|v\rangle \langle v |$ denotes the orthogonal projector onto $v\in \gH$.}
$$ h = \sum_{j = 1} ^{\infty} \lambda_j |u_j\rangle \langle u_j|$$
and define the asociated scale of Sobolev-like spaces 
\begin{equation}
\gH^s:=D(h^{s/2})=\bigg\{u=\sum_{j\geq1} \alpha_j\,u_j\ :\ \|u\|_{\gH^{s}}^2:=\sum_{j\geq1}\lambda_j^s|\alpha_j|^2<\ii\bigg\}.
\end{equation}
One can then define a finite dimensional measure on $\mathrm{span} (u_1,\ldots,u_K)$ by setting
$$d\mu_0 ^K (u) := \bigotimes_{j=1} ^K \frac{\lambda_j}{\pi} \exp\left( -\lambda_j |\langle u, u_j\rangle | ^2\right) d \langle u, u_j\rangle $$
where $d\langle u, u_j\rangle = da_j db_j$ and $a_j,b_j$ are the real and imaginary part of the scalar product. A general tightness argument for this sequence of measures allows to see that they are all cylindrical projections of a common measure living on some $\gH ^s$, where generically $s<1$: 

\begin{lemma}[\textbf{Free Gibbs measures}]\label{lem:free}\mbox{}\\
Assume that there exists $p\geq 0$ such that
\begin{equation}\label{eq:trace p}
 \tr_{\gH} \left[ h ^{-p} \right] = \sum_{j=1} ^\infty \frac{1}{\lambda_j ^p} < \infty. 
\end{equation}
Then there exists a unique measure $\mu_0$ over the space $\gH ^{1-p}$ such that, for all $K >0$, the above finite dimensional measure $\mu_{0,K}$ is the cylindrical projection of $\mu_0$ on $\mathrm{span} (u_1,\ldots,u_K)$. 
Moreover 
\begin{equation}\label{eq:DM free meas}
\gamma_0^{(k)}:=\int_{\gH^{1-p}} |u^{\otimes k}\rangle\langle u^{\otimes k}|\;d\mu_0(u) = k!\,(h^{-1})^{\otimes k} 
\end{equation}
where this is seen as an operator acting on $\bigotimes_s ^k \gH$.
\end{lemma}

Some comments:
\begin{itemize}
\item In the 1D cases with anharmonic potentials that are our main example, one can pick $p=1$, so that the measure lives on the original Hilbert space $L ^2 (\R ^d)$. In general this is not the case, and this is the main reason why we have so far been able to derive the NLS Gibbs measures only in 1D.
\item One can check that in the case of a finite interval ($a=\infty$ formally),~\eqref{eq:trace p} holds for any $p>1/2$. The measure thus lives over $\gH ^{1/2 - \eps} = H ^{1/2- \eps} ([-1,1])$, the usual $L^2$-based Sobolev space, for any $\eps$. More generally, for a Laplacian on a bounded set in $d$ dimensions, the measure lives over $H ^{1-d/2-\eps}$ for all $\eps >0$. We thus see that already in 2D, one has to consider measures over negative Sobolev spaces.
\item One might be surprised that the measure lives over a space where the energy associated with the Hamiltonian $h$ is infinite. This is however a basic fact in the theory of gaussian measures. 
\item Formula~\eqref{eq:DM free meas} indicates that although the measure itself may live over rough spaces, certain averages related to it are in fact nice compact operators over the original Hilbert space. In the cases of interest to this note, the right-hand side of~\eqref{eq:DM free meas} will actually always be trace-class. 
\end{itemize}

\subsubsection*{The interacting measure} We can now use the previous lemma to define the interacting measure. We have to check that 
$$ \FNL [u] := \frac{1}{2}\iint_{\Omega \times \Omega } |u(x)| ^2 w(x-y) |u(y)| ^2 dx dy$$
makes sense $\mu_0$-almost surely and satisfies some integrability properties w.r.t. $\mu_0$. We only need that the relative partition function $Z_r$ makes sense and is strictly positive, but in the simple case under consideration we will have much stronger properties. 

\begin{lemma}[\textbf{Interacting Gibbs measure}]\label{lem:inter}\mbox{}\\
Consider the case where $h$ is either~\eqref{eq:anharm osc} or~\eqref{eq:Lap inter}, let $p$ be the number in~\eqref{eq:trace p} and $\mu_0$ the free measure defined in Lemma~\ref{lem:free}. Assume that $w = w_1 + w_2$ where $w_1$ is a positive Radon measure and $w_2$ a positive bounded function. 

The function $ u \mapsto  \FNL [u] $ is in $L ^1 (\gH ^{1-p},d\mu_0)$. In particular 
$$\mu (du) = \frac{1}{Z_r} \exp\left( - \FNL [u] \right) \mu_0 (du)$$ 
makes sense as a probability measure over $\gH ^{1-p}$.
\end{lemma}

The proof consists of essentially two remarks:
\begin{itemize}
 \item Since we work with a positive $w$, it is sufficient to check integrability properties for $\FNL [u]$ itself to obtain $Z_r >0$. This makes computations easier.
 \item That $\FNL[u]$ is integrable follows easily from~\eqref{eq:DM free meas}. Identifying $w$ with the multiplication operator by $w(x-y)$ on the two-body Hilbert space, we have 
 \begin{align*}
 \int \FNL [u] d\mu_0 (u) &= \frac{1}{2} \int \tr \left[ w\, |u ^{\otimes 2} \rangle \langle u ^{\otimes 2} | \right] d\mu_0 (u)\\
 &= \frac{1}{4} \tr \left[ w\, h ^{-1} \otimes h ^{-1} \right],
 \end{align*}
where the trace is taken over the symmetric subspace. The expression in the right-hand side can be shown to be finite by direct estimates on the Green function~$h ^{-1}$. When $w$ is just a bounded function one can directly use the fact that $h^{-1}$ is trace-class.  
\end{itemize}

\subsubsection*{Invariance under the NLS flow}

For the particular 1D case~\eqref{eq:Lap inter}, Bourgain~\cite{Bourgain-94} proved that the NLS flow is well-defined $\mu$-almost surely and that the measure is invariant under the flow\footnote{Stricto sensu he worked under periodic boundary conditions only.}. He also considered higher order non-linearities, as well as the focusing case. Still in 1D, the harmonic oscillator case ($a=2$ in~\eqref{eq:anharm osc}, not covered by our approach) has been dealt with by Burq-Thomann-Tzvetkov~\cite{BurThoTzv-10}. Here also one gets existence of the flow $\mu$-a.s. and invariance of the measure. 

These measures thus play a special role in the theory of the NLS flow: they give a whole set of low regularity initial data for which the flow makes sense, and they are a natural invariant for the dynamics. Since the NLS equation arises as the limit of many-body quantum mechanics, it is very natural to wonder wether these objects can be related to equilibrium states of some many-body Schr\"odinger Hamiltonian. We explain that this is indeed the case in the next section.

%%%%%%%%%%%%%%%%%%%%%%%%%%%%%%%%%%%%%%%%%%
%%%%%%%%%%%%%%%%%%%%%%%%%%%%%%%%%%%%%%%%%%
\section{Derivation of the NLS-Gibbs measure on a unit interval}\label{sec:deriv}
%%%%%%%%%%%%%%%%%%%%%%%%%%%%%%%%%%%%%%%%%%
%%%%%%%%%%%%%%%%%%%%%%%%%%%%%%%%%%%%%%%%%%

In this section we turn to describing the new results proved in~\cite{LewNamRou-14d}. For simplicity we shall not present the full abstract setting but only those results that are relevant for the models based on~\eqref{eq:anharm osc} and~\eqref{eq:Lap inter}, for which~\eqref{eq:trace p} holds with $p=1$ and Lemma~\ref{lem:inter} holds true\footnote{This is refered to as ``the trace-class case'' in~\cite{LewNamRou-14d}.}. It is in fact possible to relax these assumptions and obtain a measure which lives over negative regularity spaces, whose density matrices~\eqref{eq:DM free meas} are not trace-class but only belong to a higher Schatten space. In this case the interaction operator $w$ must be very regular and cannot be of the form $w(x-y)$. One should think of it as a regularization of a physical interaction. 

The many-body object that gives rise to the nonlinear Gibbs measure is in fact the grand-canonical ensemble for bosonic particles. The limit we consider is a large temperature/mean-field limit. We shall not try to discuss the physical relevance of this limit here. The main goal is to connect nonlinear Gibbs measures to many-body equilibrium states. In view of the results recalled in Section~\ref{sec:intro} it is then very natural to expect the former to be invariant under the NLS flow.

\subsection{Grand-canonical ensemble}\label{sec:GC ens}

The grand-canonical Gibbs states live over a Hilbert space where the particle number is not fixed, but seen as a random variable. Namely, we work in the bosonic Fock space made of all possible superpositions of states with any number of particles: 
\begin{align*}
\cF &= \C \oplus \gH \oplus \gH ^{2} \oplus \ldots \oplus \gH ^{N} \oplus \ldots\\
\cF &= \C \oplus L ^2 (\Omega) \oplus L_s ^2 (\Omega ^{2})  \oplus \ldots \oplus L_s ^2 (\Omega ^{N}) \oplus \ldots 
\end{align*}
where $\Omega \subset \R$ and $\gH = L ^2 (\Om)$. Everywhere in the sequel we use the notation 
$$\gH ^k := \bigotimes_s ^k \gH$$
for the $k$-particles bosonic Hilbert space. In this context the particle number is an operator, acting on each $n$-particle sector separately, and equal to $n \1$ on the $n$-particle sector: 
$$
 \NN := \bigoplus_{n=0} ^\infty n \: \1_{\gH ^n}
$$
The mixed states over this Hilbert space are as usual the self-adjoint, positive, trace-class operators on $\cF$ with trace $1$. The pure states are the vectors of $\cF$, identified with the corresponding orthogonal projectors. For any operator $A$ on $\cF$ we call its expectation value in the state $\Gamma$ the quantity
$$ \left\langle A \right\rangle_\Gamma := \tr_{\cF}\left[ A \Gamma \right],$$
whenever this has a meaning. In particular, the expected or average particle number is 
$$ \tr_\cF \left[ \NN \Gamma \right].$$

We consider the second quantized Hamiltonian acting on $\cF$, which is the natural extension of~\eqref{eq:H_N}:
\begin{align*}
\mathbb{H}_0 &= \bigoplus_{n= 1} ^\infty \left( \sum_{j=1} ^n h_j\right) = \bigoplus_{n= 1} ^\infty \left( \sum_{j=1} ^n -\frac{d^2}{dx_j ^2} + |x_j| ^a + m \right)\\
\mathbb{W} &= \bigoplus_{n= 2} ^\infty \left( \sum_{1 \leq i<j \leq n}  w_{ij} \right) = \bigoplus_{n= 2} ^\infty \left( \sum_{1 \leq i<j \leq n}  w (x_i-x_j) \right)\\
\mathbb{H}_\lambda &= \mathbb{H}_0 + \lambda \mathbb{W} = \bigoplus_{n=1} ^\infty H_{n,\lambda}.
\end{align*}
Here $h$ is taken as in~\eqref{eq:anharm osc} or~\eqref{eq:Lap inter} (which we include as the formal $a=\infty$ case). The interaction potential $w$ is chosen as in Lemma~\ref{lem:inter}, in particular $w\geq 0$, and we introduced a coupling constant $\lambda\geq 0$. Note that this choice of $w$ is infinitesimaly form-bounded with respect to $h$, even if $w = \delta_0$, because of the 1D Sobolev embedding. The Hamiltonian thus makes sense as a self-adjoint operator.

The Gibbs state at temperature $T$ for this Hamiltonian is by definition the minimizer of the free-energy functional
\begin{equation}\label{eq:free ener GC}
\F_{\lambda} [\Gamma] = \tr_{\cF} \left[ \mathbb{H}_\lambda  \Gamma \right] + T \tr_{\cF} \left[ \Gamma \log \Gamma \right] 
\end{equation}
defined for mixed states over $\cF$. The first term is the energy, the second the von Neumann entropy. Minimizing over all states we clearly obtain 
\begin{equation}\label{eq:GC Gibbs}
 \Gamma_{\lambda,T} = \frac{1}{Z_\lambda (T)}\exp\left( - \frac1T \mathbb{H}_\lambda\right) 
\end{equation}
where the partition function $Z_\lambda (T)$ fixes the trace equal to $1$ and satisfies 
\begin{equation}\label{eq:GC part}
 F_\lambda (T) = - T \log Z_\lambda (T) 
\end{equation}
where $F_\lambda (T)$ is the minimum free energy. 

It is useful to introduce the notion of $k$-particle reduced density matrix for states oved $\cF$. For a ``diagonal'' state of the form
$$\Gamma = \bigoplus_{n=0} ^\infty G_n, \quad G_n: \gH^n \mapsto \gH ^n,$$
this is the operator 
$\Gamma ^{(k)}$ on the $k$-particles space $\gH ^k$ defined by\footnote{The partial trace is taken over the symmetric space only.}
\begin{equation}\label{eq:GC DM}
 \tr_{\gH ^k} \left[ A_k \Gamma^{(k)}\right]=\sum_{n\geq k} {n\choose k} \tr_{\gH ^n} \left[A_k \otimes_s \1 ^{\otimes (n-k)} G_n\right]
\end{equation}
for every bounded operator $A_k$ on $\gH ^k$. These are helpful for rewriting expressions involving only $k$-particle operators. In particular the number operator involves only one-body terms, so the expected particle number can be computed as
\begin{equation}\label{eq:exp number}
 \tr_{\cF} \left[ \NN \Gamma \right] = \tr_{\gH} \left[ \Gamma ^{(1)}\right].  
\end{equation}
The energy involves only two-body terms, and we have: 
\begin{equation}\label{eq:ener dens mat} 
\tr_{\cF} \left[ \mathbb{H}_\lambda  \Gamma \right] = \tr_{\gH ^1} \left[ h \: \Gamma ^{(1)}\right] + \lambda \tr_{\gH ^2} \left[ w \: \Gamma ^{(2)}\right].
\end{equation}
The limit we shall consider is 
\begin{equation}\label{eq:limit}
T \to \infty,\quad \lambda T \to 1. 
\end{equation}
The rationale here is that to see a non-trivial effect of the temperature in a bosonic system, $T$ should be rather large, because of the Bose-Einstein condensation phenomenon we alluded to before. Then one can easily get convinced that the expected particle number behaves as $T$ in this limit. For example, consider the $\lambda = 0$ case. The one-particle density matrix can be computed explicitly:
$$ \Gamma_{0,T}^{(1)}=\frac{1}{e^{h/T}-1}.$$
After dividing by $T$, this clearly converges to the $k=1$ case of~\eqref{eq:DM free meas}. Since we assume the latter object to have a finite trace, in view of~\eqref{eq:exp number}, it is clear that the expected particle number of the free Gibbs state grows like $T$. Taking the coupling constant $\lambda \sim 1 / T$ is thus the equivalent, in the grand-canonical setting, of the canonical mean-field limit discussed in Section~\ref{sec:intro} where the coupling constant scaled as the inverse of the particle number. Note that in case $p>1$ in~\eqref{eq:trace p}, this argument is no longer valid, and the particle number will grow much faster than $T$. This makes this case much more difficult to deal with, which is why it is not included in this note and we refer to the original paper.

\subsection{Main theorem}\label{sec:main}

Let us now state the main result of this note, proved in~\cite{LewNamRou-14d}. It concerns the limit~\eqref{eq:limit} of the previously defined grand-canonical ensemble. One should not expect a convergence of the Gibbs states themselves since their expected particle number diverges. Neither should one hope for an estimate in some norm, for well-known reasons that we shall not elaborate on here. The correct objects to look at are the $k$-body density matrices with $k$ much smaller than the particle number (in particular, $k$ fixed will do). The results on ground states and dynamics we have mentioned in Section~\ref{sec:intro} are all of the form of estimates on such objects, and so is the following theorem on positive temperature equilibrium states:

\begin{theorem}[\textbf{Derivation of nonlinear Gibbs measures}]\label{thm:main}\mbox{}\\
Let $h$ be given by~\eqref{eq:anharm osc} or~\eqref{eq:Lap inter}, and $w$ satisfy the assumptions of Lemma~\ref{lem:inter}. Let $\mu$ be the Gibbs measure defined therein. For every fixed $k\geq1$ we have
\begin{equation}\label{eq:result DM}
\frac{k!}{T^k}\Gamma_{\lambda,T}^{(k)} \to \int_\gH |u^{\otimes k}\rangle\langle u^{\otimes k}|\,d\mu(u)
\end{equation}
strongly in the trace-class norm, in the limit~\eqref{eq:limit}.  
\end{theorem}

Some comments:
\begin{itemize}
\item When $L ^2 (\Om)$ is replaced by a finite dimensional Hilbert space, canonical versions of this theorem have been proved before,  see~\cite{Gottlieb-05,Knowles-thesis} and~\cite[Appendix~B]{Rougerie-LMU}.
 \item The collection of all the ``moments'' of $\mu$ for $k=1,2 \ldots$ appearing in the right-hand side of~\eqref{eq:result DM} characterizes  $\mu$ uniquely. This measure is thus the natural, unique, limit object for the grand-canonical ensemble.
 \item Roughly speaking, ~\eqref{eq:result DM} is a way of making rigorous the approximation 
 \begin{equation}\label{eq:trial}
 \Gamma_{\lambda,T} \approx \int \big| \xi (\sqrt{T}u) \big\rangle \big\langle \xi (\sqrt{T} u ) \big|\, d\mu (u) 
 \end{equation}
where $\xi(v)$ is the coherent state built on $v$ :
\begin{equation}\label{eq:coherent state}
 \xi(v) = e ^{-\norm{v} ^2/2} \bigoplus_{k\geq 0} \frac{v ^{\otimes k}}{\sqrt{k!}}. 
\end{equation}
\end{itemize}

\subsection{The variational formulation}\label{sec:vari}

The proof of Theorem~\ref{thm:main} is variational. The convergence of density matrices is obtained as a corollary of estimates on the minimum free energy, or equivalently, the partition function. An important difficulty is however that $Z_\lambda (T)$ is difficult to control directly, in particular because even at $\lambda = 0$ this quantity diverges very fast when $T\to \infty$. Our strategy is therefore to instead deal with the difference between the free energies at $\lambda = 0$ and $\lambda >0$, that is, to count everything relatively to the free Gibbs state. This is possible because, as one easily realizes, $\Gamma_{\lambda,T}$ also minimizes the relative free-energy functional
\begin{align}\label{eq:rel free ener}
\F_{\rm rel} [\Gamma] &= \lambda \tr_{\cF} \left[ \mathbb{W} \: \Gamma \right] + T \tr_{\cF} \left[ \Gamma \left(\log \Gamma - \log \Gamma_{0,T} \right)\right] \nonumber\\
&=\lambda \tr_{\gH ^2} \left[ w \: \Gamma ^{(2)}\right] + T \tr_{\cF} \left[ \Gamma \left(\log \Gamma - \log \Gamma_{0,T} \right)\right]
\end{align}
amongst all mixed states. The second term in the above is the (quantum) relative entropy of $\Gamma$ with respect to $\Gamma_{0,T}$, which is always positive. The infimum of the above functional equals
$$ F_\lambda (T) - F_0 (T) = T \log \frac{Z_0 (T)}{Z_\lambda (T)}.$$
Studying the minimization of~\eqref{eq:rel free ener} thus allows for a direct access to the ratio of the interacting and non-interacting partition functions, which will converge to some limit although the partition functions themselves diverge. We obtain Theorem~\ref{thm:main} as a corollary of the estimate

\begin{theorem}[\textbf{Asymptotics of the free-energy}]\label{thm:free ener}\mbox{}\\
Let $Z_r$ be the relative partition function for the Gibbs measure $\mu$, defined in Lemma~\ref{lem:inter}, 
$$ Z_r  = \int \exp\left(- \FNL [u]\right) d\mu_0 (u)$$
with $\mu_0$ the free Gibbs measure. In the limit~\eqref{eq:limit} we have 
$$\frac{F_\lambda (T) - F_0 (T)}{T} \to  -\log Z_r. $$
\end{theorem}
 
In order to complete the variational framework of the proof, one characterizes $-\log Z_r$ as the infimum of some variational problem of which $\mu$ is the solution. Indeed, one easily sees that $\mu$ minimizes 
\begin{align}\label{eq:class rel ener}
\F_{\rm cl} [\nu] &= \int \FNL [u] d\nu (u) + \int \frac{d\nu}{d\mu_0} \log \left( \frac{d\nu}{d\mu_0}\right) d\mu_0\nonumber\\
&= \int \frac{1}{2} \tr_{\gH ^2} \left[w\, |u ^{\otimes 2} \rangle \langle u ^{\otimes 2}| \right] d\nu (u) + \int \frac{d\nu}{d\mu_0} \log \left( \frac{d\nu}{d\mu_0}\right) d\mu_0
\end{align}
amongst all $\mu_0$-absolutely continuous measures $\nu$ on the one-body Hilbert space $\gH$. The second term is the classical relative entropy of $\nu$ relative to $\mu_0$.
 
The semi-classical nature of Theorems~\ref{thm:main} and~\ref{thm:free ener} is now apparent.  We have to relate the minimization problems for the quantum and classical free energy functionals~\eqref{eq:rel free ener} and~\eqref{eq:class rel ener}. This means that we need to understand how to describe quantum objects, operators on a Hilbert space, in terms of classical objects, probability measures. This requires specific tools, and we shall review some of these in the next section. 

Since we work on a variational problem, we can provide separately an upper and a lower bound on the free energy. The former essentially follows from the evaluation of the free energy of a trial state of the form~\eqref{eq:trial}. Using some projected states in an appropriate way, this can be done by adapting well-known tools of finite dimensional semi-classical analysis. The lower bound is as usual the tougher part, and we will try to give (some of) the main ideas in the sequel.

%%%%%%%%%%%%%%%%%%%%%%%%%%%%%%%%%%%%%%%%%%
%%%%%%%%%%%%%%%%%%%%%%%%%%%%%%%%%%%%%%%%%%
\section{Some tools of the proof}\label{sec:proof}
%%%%%%%%%%%%%%%%%%%%%%%%%%%%%%%%%%%%%%%%%%
%%%%%%%%%%%%%%%%%%%%%%%%%%%%%%%%%%%%%%%%%%

The first point is to characterize the (limits of) quantum states over the Fock space in terms of classical objects, i.e. probability measures over the one-body Hilbert space. This is the object of the so-called quantum de Finetti theorem, that we discuss first. It will be rather clear from this kind of results how one can deal with the interaction energy term in~\eqref{eq:rel free ener}. The relative entropy term requires new specific tools, discussed in a second subsection.

\subsection{Quantum de Finetti measures}\label{sec:QdeF}

The main message of the quantum de Finetti theorem is informally that, ``For many practical purposes, any bosonic state looks like a convex combination of product states''. A little bit more precisely, if one picks a bosonic $N$-body state, with $N$ large, and look at its $k$-body density matrix, for $k$ small, one obtains ``almost'' a convex combination of product states.

There is a vast literature devoted to rigorous statements and proofs of these claims. We shall not review it here in details, see~\cite{Ammari-HDR,Rougerie-cdf,Rougerie-LMU}. For the purpose of this note, we rely on recent approaches and versions of the theorem due mainly to Ammari-Nier and Lewin-Nam-Rougerie~\cite{AmmNie-08,AmmNie-09,AmmNie-15,LewNamRou-14}. To deal with the case where $p>1$ in~\eqref{eq:trace p}, some improvements are necessary, provided in~\cite{LewNamRou-14d}, but for the proof of the previous theorems, the following result of~\cite{AmmNie-08} is sufficient

\begin{theorem}[\textbf{Grand-canonical quantum de Finetti theorem}]\label{thm:deF}\mbox{}\\
Let $0\leq \Gamma_n$ be a sequence of states on the Fock space $\cF(\gH)$. Assume that there exists a sequence $0<\epsilon_n\to0$ such that, for all $k\in \N $
\begin{equation}
\epsilon_n ^k  \tr_{\gH ^{k}} \Big[ \Gamma_n ^{(k)}\Big]\leq C_k \ <\infty.
\label{eq:bound_DM}
\end{equation}
Then, there exists a unique Borel probability measure $\nu$ on $\gH$ such that, along a subsequence,  
\begin{equation}
k!(\epsilon_{n})^k\,\Gamma_{n}^{(k)} \wto \int_{\gH} |u^{\otimes k}\rangle \langle u^{\otimes k}| \,d\nu(u)
\label{eq:weak_limit_DM}
\end{equation}
weakly in the trace-class for every integer $k$.
\end{theorem}

Some comments:
\begin{itemize}
 \item The measure $\nu$ is called the de Finetti measure of the sequence $\Gamma_n$. It is the classical object we are after. Its existence is a rather general fact, based mostly on the bosonic symmetry of the states under consideration. Theorems~\ref{thm:main} and~\ref{thm:free ener} can be rephrased as saying that the de Finetti measure of the sequence $\Gamma_{\lambda,T}$ is the nonlinear Gibbs measure $\mu$. 
  \item Here $\epsilon_n$ should be thought of as a semiclassical parameter essentially forcing the particle number to behave as $\eps_n ^{-1}$. To apply this result in our context, we need to check that~\eqref{eq:bound_DM} holds, with $\epsilon_n = T ^{-1}$ and $\Gamma = \Gamma_{\lambda,T}$. In the context of this note, this is not too difficult because the particle number is well under control. Bounds on the expectation of $\NN ^k$ in some state naturally translate into trace-class bounds on the $k$-body density matrix.
\item Since we here get convergence of density matrices, a lower bound to the interaction energy term in~\eqref{eq:rel free ener} follows from this theorem.
\end{itemize}

\subsection{Berezin-Lieb inequality for the relative entropy}\label{sec:BL ineq}

There remains to understand how to estimate the relative entropy term in~\eqref{eq:rel free ener}. One should keep the following in mind:
\begin{itemize}
 \item Contrarily to the energy, it is not expressed in terms of reduced density matrices only. It genuinely depends of the full state.
 \item One should not ``undo'' the relative entropy to consider only the von Neumnann entropies of $\Gamma_{0,T}$ and $\Gamma_{\lambda,T}$ directly, since we do not expect a good control on those.
\end{itemize}

The main tool to link quantum entropy terms to their semiclassical counterparts are the so-called Berezin-Lieb inequalities~\cite{Berezin-72,Lieb-73b,Simon-80}. Since we work in infinite dimensions and we cannot undo the relative entropy to rely on known inequalities, we need the following new result:

\begin{theorem}[\textbf{Berezin-Lieb inequality for the relative entropy}]\label{thm:BL}\mbox{}\\
Let $\epsilon_n \to 0$ and $\{\Gamma_n\}$, $\{\Gamma_n'\}$ be two sequences of states  satisfying the assumptions of Theorem~\ref{thm:deF}. Let $\nu$ and $\nu'$ be their de Finetti measures. Then
\begin{equation}\label{eq:BL ineq}
\liminf_{n\to \infty} \tr_{\cF} \left[ \Gamma_n \left( \log \Gamma_n - \log \Gamma_n' \right)\right] \ge \int \frac{d\nu}{d\nu'} \log \left(\frac{d\nu}{d\nu'}\right) d\nu'. 
\end{equation}
\end{theorem}

Some comments:

\begin{itemize}
\item Clearly, this will give the desired lower bound to the relative entropy term in~\eqref{eq:rel free ener} if we can prove first that the de Finetti measure of the free Gibbs state $\Gamma_{0,T}$ is given by $\mu_0$. Since all the density matrices of $\Gamma_{0,T}$ can be explicitly computed, this is easily shown by direct inspection.
\item For the proof of this result, we rely heavily on a constructive approach to the proof of Theorem~\ref{thm:deF}. Indeed, not only the existence of the measure is crucial, but also the precise way one can approach it by explicit functions of the states. 
\item The link with semi-classics proceeds by seeing the de Finetti measure both as a lower symbol/covariant symbol/Husimi function/anti-Wick quantization and as an approximate upper symbol/contravariant symbol/Wigner measure/Wick quantization based on a coherent state decomposition (made of projectors onto states of the form~\eqref{eq:coherent state}).
\item We also rely on deep properties of the quantum relative entropy: joint convexity and monotonicity under two-positive trace preserving maps~\cite{Wehrl-78,OhyPet-93,Carlen-10}. These are in fact equivalent to the strong subadditivity of the quantum entropy, a crucial property proved by Lieb and Ruskai~\cite{LieRus-73a,LieRus-73b}.
\end{itemize}

\bigskip

\noindent \textbf{Acknowledgment.} The research summarized here has received funding from the \emph{European Research Council} (M.L., ERC Grant Agreement MNIQS 258023), the \emph{People Programme / Marie Curie Actions} (P.T.N., REA Grant Agreement 291734) and from the \emph{ANR} (Projects NoNAP ANR-10-BLAN-0101 \& Mathostaq ANR-13-JS01-0005-01).

%%%%%%%%%%%%%%%%%%%%%%%%%%%%%%%%%%%%%%%%%%
%%%%%%%%%%%%%%%%%%%%%%%%%%%%%%%%%%%%%%%%%%
% \bibliographystyle{hacm}
% \bibliography{biblio_NR_Jun15}

\begin{thebibliography}{10}

\bibitem{Ammari-HDR}
{\sc Ammari, Z.}
\newblock Systèmes hamiltoniens en théorie quantique des champs : dynamique
  asymptotique et limite classique.
\newblock Habilitation {\`a} Diriger des Recherches, University of Rennes I,
  February 2013.

\bibitem{AmmNie-08}
{\sc Ammari, Z., and Nier, F.}
\newblock Mean field limit for bosons and infinite dimensional phase-space
  analysis.
\newblock {\em Ann. Henri Poincar\'e 9\/} (2008), 1503--1574.

\bibitem{AmmNie-09}
{\sc Ammari, Z., and Nier, F.}
\newblock Mean field limit for bosons and propagation of {W}igner measures.
\newblock {\em J. Math. Phys. 50}, 4 (2009), 042107.

\bibitem{AmmNie-15}
{\sc Ammari, Z., and Nier, F.}
\newblock {Mean field propagation of infinite dimensional Wigner measures with
  a singular two-body interaction potential}.
\newblock {\em Ann. Sc. Norm. Sup. Pisa.\/} (2015).

\bibitem{BenPorSch-15}
{\sc {Benedikter}, N., {Porta}, M., and {Schlein}, B.}
\newblock {Effective Evolution Equations from Quantum Dynamics}.
\newblock {\em ArXiv e-prints\/} (May 2015), 1502.02498.

\bibitem{BenLie-83}
{\sc {Benguria}, R., and {Lieb}, E.~H.}
\newblock {Proof of the Stability of Highly Negative Ions in the Absence of the
  Pauli Principle}.
\newblock {\em Phys. Rev. Lett. 50\/} (May 1983), 1771--1774.

\bibitem{Berezin-72}
{\sc Berezin, F.~A.}
\newblock Convex functions of operators.
\newblock {\em Mat. Sb. (N.S.) 88(130)\/} (1972), 268--276.

\bibitem{Bogachev}
{\sc Bogachev, V.~I.}
\newblock {\em Gaussian measures}.
\newblock No.~62. American Mathematical Soc., 1998.

\bibitem{Bourgain-94}
{\sc Bourgain, J.}
\newblock Periodic nonlinear {S}chr\"odinger equation and invariant measures.
\newblock {\em Comm. Math. Phys. 166}, 1 (1994), 1--26.

\bibitem{Bourgain-96}
{\sc Bourgain, J.}
\newblock Invariant measures for the 2d-defocusing nonlinear {S}chr\"odinger
  equation.
\newblock {\em Comm. Math. Phys. 176\/} (1996), 421--445.

\bibitem{Bourgain-97}
{\sc Bourgain, J.}
\newblock {Invariant measures for the Gross-Piatevskii equation}.
\newblock {\em Journal de Mathématiques Pures et Appliquées 76}, 8 (1997),
  649--02.

\bibitem{BurThoTzv-10}
{\sc {Burq}, N., {Thomann}, L., and {Tzvetkov}, N.}
\newblock {Long time dynamics for the one dimensional non linear Schr\"odinger
  equation}.
\newblock {\em Ann. Inst. Fourier. 63\/} (2013), 2137--2198.

\bibitem{BurTzv-08}
{\sc Burq, N., and Tzvetkov, N.}
\newblock Random data {C}auchy theory for supercritical wave equations. {I}.
  {L}ocal theory.
\newblock {\em Invent. Math. 173}, 3 (2008), 449--475.

\bibitem{CacSuz-14}
{\sc Cacciafesta, F., and {de Suzzoni}, A.-S.}
\newblock Invariant measure for the {S}chr\"odinger equation on the real line.
\newblock {\em arXiv eprints\/} (2014), 1405.5107.

\bibitem{Carlen-10}
{\sc Carlen, E.}
\newblock Trace inequalities and quantum entropy: an introductory course.
\newblock In {\em Entropy and the Quantum\/} (2010), R.~Sims and D.~Ueltschi,
  Eds., vol.~529 of {\em Contemporary Mathematics}, American Mathematical
  Society, pp.~73--140.
\newblock Arizona School of Analysis with Applications, March 16-20, 2009,
  University of Arizona.

\bibitem{DerGer-13}
{\sc Derezi{\'n}ski, J., and G{\'e}rard, C.}
\newblock {\em {Mathematics of Quantization and Quantum Fields}}.
\newblock Cambridge University Press, Cambridge, 2013.

\bibitem{ErdSchYau-09}
{\sc Erd{\H{o}}s, L., Schlein, B., and Yau, H.-T.}
\newblock Rigorous derivation of the {G}ross-{P}itaevskii equation with a large
  interaction potential.
\newblock {\em J. Amer. Math. Soc. 22}, 4 (2009), 1099--1156.

\bibitem{FanSpoVer-80}
{\sc Fannes, M., Spohn, H., and Verbeure, A.}
\newblock Equilibrium states for mean field models.
\newblock {\em J. Math. Phys. 21}, 2 (1980), 355--358.

\bibitem{FroKnoSch-09}
{\sc Fr{\"o}hlich, J., Knowles, A., and Schwarz, S.}
\newblock On the mean-field limit of bosons with {C}oulomb two-body
  interaction.
\newblock {\em Commun. Math. Phys. 288}, 3 (2009), 1023--1059.

\bibitem{GinVel-79}
{\sc Ginibre, J., and Velo, G.}
\newblock The classical field limit of scattering theory for nonrelativistic
  many-boson systems. {I}.
\newblock {\em Commun. Math. Phys. 66}, 1 (1979), 37--76.

\bibitem{GliJaf-87}
{\sc Glimm, J., and Jaffe, A.}
\newblock {\em Quantum Physics: A Functional Integral Point of View}.
\newblock Springer-Verlag, 1987.

\bibitem{Golse-13}
{\sc {Golse}, F.}
\newblock {On the Dynamics of Large Particle Systems in the Mean Field Limit}.
\newblock {\em ArXiv e-prints 1301.5494\/} (Jan. 2013), 1301.5494.
\newblock Lecture notes for a course at the NDNS+ Applied Dynamical Systems
  Summer School "Macroscopic and large scale phenomena", Universiteit Twente,
  Enschede (The Netherlands).

\bibitem{Gottlieb-05}
{\sc Gottlieb, A.~D.}
\newblock Examples of bosonic de {F}inetti states over finite dimensional
  {H}ilbert spaces.
\newblock {\em J. Stat. Phys. 121}, 3-4 (2005), 497--509.

\bibitem{Hepp-74}
{\sc Hepp, K.}
\newblock The classical limit for quantum mechanical correlation functions.
\newblock {\em Comm. Math. Phys. 35}, 4 (1974), 265--277.

\bibitem{Knowles-thesis}
{\sc Knowles, A.}
\newblock Limiting dynamics in large quantum systems.
\newblock Doctoral thesis, ETH Z\"urich.

\bibitem{LebRosSpe-88}
{\sc Lebowitz, J.~L., Rose, H.~A., and Speer, E.~R.}
\newblock Statistical mechanics of the nonlinear {S}chr\"odinger equation.
\newblock {\em J. Statist. Phys. 50}, 3-4 (1988), 657--687.

\bibitem{LewNamRou-14d}
{\sc Lewin, M., Nam, P., and Rougerie, N.}
\newblock Derivation of nonlinear {G}ibbs measures from many-body quantum
  mechanics.
\newblock {\em arXiv eprints\/} (2014), 1410.0335.

\bibitem{LewNamRou-14}
{\sc Lewin, M., Nam, P.~T., and Rougerie, N.}
\newblock Derivation of {H}artree's theory for generic mean-field {B}ose gases.
\newblock {\em Adv. Math. 254\/} (March 2014), 570--621, 1303.0981.

\bibitem{Lieb-73b}
{\sc Lieb, E.~H.}
\newblock The classical limit of quantum spin systems.
\newblock {\em Comm. Math. Phys. 31\/} (1973), 327--340.

\bibitem{LieRus-73a}
{\sc Lieb, E.~H., and Ruskai, M.~B.}
\newblock A fundamental property of quantum-mechanical entropy.
\newblock {\em Phys. Rev. Lett. 30\/} (1973), 434--436.

\bibitem{LieRus-73b}
{\sc Lieb, E.~H., and Ruskai, M.~B.}
\newblock Proof of the strong subadditivity of quantum-mechanical entropy.
\newblock {\em J. Math. Phys. 14\/} (1973), 1938--1941.
\newblock With an appendix by B. Simon.

\bibitem{LieSei-06}
{\sc Lieb, E.~H., and Seiringer, R.}
\newblock Derivation of the {G}ross-{P}itaevskii equation for rotating {B}ose
  gases.
\newblock {\em Commun. Math. Phys. 264}, 2 (2006), 505--537.

\bibitem{LieSeiSolYng-05}
{\sc Lieb, E.~H., Seiringer, R., Solovej, J.~P., and Yngvason, J.}
\newblock {\em The mathematics of the {B}ose gas and its condensation}.
\newblock Oberwolfach {S}eminars. Birkh{\"a}user, 2005.

\bibitem{LieYau-87}
{\sc Lieb, E.~H., and Yau, H.-T.}
\newblock The {C}handrasekhar theory of stellar collapse as the limit of
  quantum mechanics.
\newblock {\em Commun. Math. Phys. 112}, 1 (1987), 147--174.

\bibitem{LorHirBet-11}
{\sc L{\"o}rinczi, J., Hiroshima, F., and Betz, V.}
\newblock {\em {Feynman-Kac-Type Theorems and Gibbs Measures on Path Space:
  With Applications to Rigorous Quantum Field Theory}}.
\newblock Gruyter - de Gruyter Studies in Mathematics. Walter de Gruyter GmbH
  \& Company KG, 2011.

\bibitem{NamRouSei-15}
{\sc Nam, P.~T., Rougerie, N., and Seiringer, R.}
\newblock {Ground states of large Bose systems: The Gross-Pitaevskii limit
  revisited}.
\newblock {\em preprint arXiv\/} (2015), arXiv:1503.07061.

\bibitem{OhyPet-93}
{\sc Ohya, M., and Petz, D.}
\newblock {\em Quantum entropy and its use}.
\newblock Texts and Monographs in Physics. Springer-Verlag, Berlin, 1993.

\bibitem{Pickl-11}
{\sc Pickl, P.}
\newblock A simple derivation of mean-field limits for quantum systems.
\newblock {\em Lett. Math. Phys. 97}, 2 (2011), 151--164.

\bibitem{RagWer-89}
{\sc Raggio, G.~A., and Werner, R.~F.}
\newblock Quantum statistical mechanics of general mean field systems.
\newblock {\em Helv. Phys. Acta 62}, 8 (1989), 980--1003.

\bibitem{RodSch-09}
{\sc Rodnianski, I., and Schlein, B.}
\newblock Quantum fluctuations and rate of convergence towards mean field
  dynamics.
\newblock {\em Commun. Math. Phys. 291}, 1 (2009), 31--61.

\bibitem{Rougerie-LMU}
{\sc Rougerie, N.}
\newblock De finetti theorems, mean-field limits and bose-einstein
  condensation.
\newblock arXiv:1506.05263, 2014.
\newblock Lecture Notes for a course at LMU, Munich.

\bibitem{Rougerie-cdf}
{\sc Rougerie, N.}
\newblock Théorèmes de de {F}inetti, limites de champ moyen et condensation de
  {B}ose-{E}instein.
\newblock Lecture notes for a cours Peccot, 2014.

\bibitem{Schlein-08}
{\sc Schlein, B.}
\newblock Derivation of effective evolution equations from microscopic quantum
  dynamics.
\newblock {\em arXiv eprints\/} (2008), 0807.4307.
\newblock Lecture Notes for a course at ETH Zurich.

\bibitem{Simon-74}
{\sc Simon, B.}
\newblock {\em The {$P(\phi )_{2}$} {E}uclidean (quantum) field theory}.
\newblock Princeton University Press, Princeton, N.J., 1974.
\newblock Princeton Series in Physics.

\bibitem{Simon-80}
{\sc Simon, B.}
\newblock The classical limit of quantum partition functions.
\newblock {\em Comm. Math. Phys. 71}, 3 (1980), 247--276.

\bibitem{Skorokhod-74}
{\sc Skorokhod, A.}
\newblock {\em Integration in {H}ilbert space}.
\newblock Ergebnisse der Mathematik und ihrer Grenzgebiete. Springer-Verlag,
  1974.

\bibitem{Spohn-80}
{\sc Spohn, H.}
\newblock Kinetic equations from {H}amiltonian dynamics: {M}arkovian limits.
\newblock {\em Rev. Modern Phys. 52}, 3 (1980), 569--615.

\bibitem{Summers-12}
{\sc {Summers}, S.~J.}
\newblock {A Perspective on Constructive Quantum Field Theory}.
\newblock {\em ArXiv e-prints\/} (Mar. 2012), 1203.3991.

\bibitem{ThoTzv-10}
{\sc Thomann, L., and Tzvetkov, N.}
\newblock Gibbs measure for the periodic derivative nonlinear schrödinger
  equation.
\newblock {\em Nonlinearity 23}, 11 (2010), 2771.

\bibitem{Tzvetkov-08}
{\sc Tzvetkov, N.}
\newblock Invariant measures for the defocusing nonlinear {S}chr\"odinger
  equation.
\newblock {\em Ann. Inst. Fourier (Grenoble) 58}, 7 (2008), 2543--2604.

\bibitem{VelWig-73}
{\sc Velo, G., and Wightman, A.}, Eds.
\newblock {\em {Constructive quantum field theory: The 1973 Ettore Majorana
  international school of mathematical physics}}.
\newblock Lecture notes in physics. Springer-Verlag, 1973.

\bibitem{Wehrl-78}
{\sc Wehrl, A.}
\newblock General properties of entropy.
\newblock {\em Rev. Modern Phys. 50}, 2 (1978), 221--260.

\end{thebibliography}

\end{document}